\documentclass[
    ,final            
  ]
  {aipproc}

\layoutstyle{6x9}

\begin{document}

\title{Probability distributions and Gleason's Theorem}


\author{Sebastian Rieder}{
  address={Institut f\"ur Theoretische Physik, University of Technology Vienna,
Wiedner Hauptstra\ss e 8-10/136, A-1040 Vienna, Austria}
}
\author{Karl Svozil}{
  address={Institut f\"ur Theoretische Physik, University of Technology Vienna,
Wiedner Hauptstra\ss e 8-10/136, A-1040 Vienna, Austria}
}

\begin{abstract}
We discuss concrete examples for frame functions and their associated density operators, as well as for non-Gleason type probability measures.
\end{abstract}

\maketitle

\section{Introduction}

The purpose of this paper is to deepen the understanding of the importance of Gleason's theorem in
connection with probability distributions and the density matrix formalism,
as well as the discussion of non-Gleason type probabilities.
Before we present example calculations from frame functions via probability
distributions to density matrices and  {\it vice versa} we review some theory,
starting with a short characterization of density matrices and proceeding
with a discussion of Gleason theorem.

\subsection{Quantum states}

In what follows, only real Hilbert spaces will be considered.
Within von Neumann's  Hilbert space framework
\cite{v-neumann-49}, the state of a given quantized system can always be expressed by a density operator.
A system in a pure state  $\vert \psi \rangle \equiv x$ is fully
characterized by a state vector  $x$.  Its density operator is the projector
$\rho = \vert \psi \rangle  \langle  \psi \vert \equiv
x^T\otimes x$, where the superscript $T$ indicates transposition, and
$\otimes$ stands for the dyadic or tensor product.
Generally, the density operator of the mixed state is defined as a convex combination
$ \rho =
\sum_{i=1}^m p_{i} \vert \psi_{i}\rangle \langle \psi_{i} \vert$
of normalized pure state vectors $\{\vert \psi_i\rangle \}$, where
$0\ge p_{i}\ge 1$ and $\sum_{i=1}^m p_{i}=1$.
The density operator is a positiv definite Hermitian operator with unit
trace.
According to the {\em Born rule}, the expectation value
$\langle A\rangle$  of an observable $A$ is the trace of $\rho A$; i.e.,
$\langle A\rangle ={\rm tr}(\rho  A)$.
In particular, if $A$ is a projector $E$ corresponding to an elementary yes-no proposition
{\it ``the system has property Q,''} then $\langle E\rangle ={\rm tr}(\rho  E)$ corresponds
to the probability of that property $Q$ if the system is in state $\rho$.
The equations $\rho ^{2} = \rho$ and ${\rm tr}(\rho  ^{2})=1$
are only valid for pure states, because $\rho $ is not idempotent for mixed states.

\subsection{Gleason's theorem}

One way of interpreting Gleason's theorem
\cite{Gleason,r:dvur-93,c-k-m,peres,hru-pit-2003,rich-bridge}
is to view it as a derivation of the Born rule
from fundamental assumptions about quantum probabilities, guided by
quantum theory, in order to assign consistent and unique probabilities to all
possible measurement outcomes.
With these provisos, Gleason proved that there is no alternative to
the Born rule for Hilbert spaces of dimension greater than two.

Before we introduce Gleason's Theorem, we have to define a measure
$\varepsilon $ on our real or complex Hilbert space. A measure is a mapping
which assigns a nonnegative real number $\varepsilon (E)$  to each projector $E$
such that, if $E_i$ are mutually orthogonal, then the measure of
$\sigma =\sum_i E_i$ has to
satisfy the (sub-)additivity property
$\varepsilon (\sigma )=\sum_i  \varepsilon (E_i)$.
Any such measure is determined by its values on the one dimensional projections.
Consider a unit vector $x$ and the associated one dimensional projector $E_x= x^T\otimes x$,
then the measure $\varepsilon $ is determined by the real-valued positive function
$f(x)=\varepsilon (E_x)$ on the unit sphere. The weight $W$ of the
function $f$  is defined as the measure of the identity projection;
i.e., $W=\varepsilon ({\bf 1})$. Then the
function satisfies $\sum_i f(e_i)=W$
for each orthonormal basis $\{e_i\}$. These functions are called frame functions of
weight $W$.

For quantum probability theory, the value of the weight W is necessarily $1$.
The physical meaning of a frame function $f(x)$ is
the probability of the proposition associated with the projector $E_x$
for a given quantum system in a state associated with $f$.
Gleason's Theorem can be stated as follows:
Let $H$ be a Hilbert space of dimension greater than two. Let $f$ be
a frame function.
Then there exists a unique density operator (i.e., a positive operator of trace class) $\rho $ on $H$
such that
$f(x)= {\rm tr} (\rho E_x)
=
\sum_i \langle e_i\vert \rho E_x \vert e_i\rangle
=
\sum_{i,j} \langle e_i \vert \rho \vert e_j\rangle \langle e_j \vert E_x \vert e_i\rangle
=
\sum_{i,j} \langle e_i \vert \rho \vert e_j\rangle \langle e_j \vert x\rangle \langle x \vert e_i\rangle
=
\langle \rho x \vert x\rangle $ is a quadratic form  of $x$ for all
projectors $E_x$ associated with elementary yes-no propositions.

Let us summarize what could be called the essence of Gleason's Theorem.
Roughly speaking, from the assumption
of {\em quasi-classical probabilities} for {\em co-measurable} events associated with {\em commuting} operators
(projectors)
follows the Born rule of quantum probabilities.
In other words, if we want a (positive) probability measure to be totally
(sub-)additive on all subspaces or projections (of a Hilbert space of finite dimension
greater than two) which are co-measurable, then the only possibility is the density matrix formalism.
Stated differently, the Born quantum probability rule follows from
elementary assumptions about quantum mechanics, the
representation of observables by self-adjoint operators in Hilbert spaces,
and from the {\em quasi-classical} consistent assignment of probabilities for {\em compatible} measurement outcomes.

Note that classically the domain for which the frame (probability) function is (sub-)additive
extends over all observables. Moreover, any classical probability distribution can be
written as the convex sum of a set of singular, two-valued measures.
The Kochen-Specker \cite{kochen1} theorem states that, for Hilbert spaces of dimension greater than two,
no two-valued measure and thus no such represention exists.

\section{The easy part: calculation of frame functions from density operators}

To calculate a frame function from a given density operator, one has to
use the formula $f(x)=\langle \rho x \vert  x\rangle $ explicitly, as given by Gleason's Theorem
in a straightforward calculation.

\subsection{Pure states}

The most elementary example is the state corresponding to
$\vert \Psi \rangle  \equiv (1,0,0)$.
Since the associated density operator is the projector  corresponding to the diagonal matrix
$\rho_\Psi ={\rm diag} (1,0,0)$,
with $x=(x_1,x_2,x_3)$, the frame function ist just $f_{\Psi}(x)=x_1^2$.

As another example,
consider the four Bell basis states
$\vert \Psi^{\pm} \rangle  \equiv (1/\sqrt{2}) (1,0,0,\pm 1)$
and
$\vert \Phi^{\pm} \rangle  \equiv (1/\sqrt{2}) (0,1,\pm 1,0)$.

The associated density
matrices are given by
$$
\begin{array}{ccc}
\rho_{\Psi^{\pm}} =
{1\over \sqrt{2}}
\left(
\begin{array}{c}
1\\
0\\
0\\
\pm 1
\end{array}
\right)
\otimes
{1\over \sqrt{2}} (1,0,0,\pm 1)
&=&
{1\over 2}
\left(
\begin{array}{cccc}
1&0&0&\pm 1\\
0&0&0&0\\
0&0&0&0\\
\pm 1&0&0&1
\end{array}
\right),
\\
\rho_{\Phi^{\pm}} =
{1\over \sqrt{2}}
\left(
\begin{array}{c}
0\\
1\\
\pm 1\\
0
\end{array}
\right)
\otimes
{1\over \sqrt{2}} (0,1,\pm 1,0)
&=&
{1\over 2}
\left(
\begin{array}{cccc}
0&0&0&0\\
0&1&\pm 1&0\\
0&\pm 1&1&0\\
0&0&0&0
\end{array}
\right).
\end{array}
$$
For $x=(x_1,x_2,x_3,x_4)$, the associated frame functions turn out to be
$$
\begin{array}{ccc}
f_{\Psi^{\pm}}(x) &=& {1\over 2}\left( x_1\pm x_4\right)^2,\\
f_{\Phi^{\pm}}(x) &=& {1\over 2}\left( x_2\pm x_3\right)^2.
\end{array}
$$

\subsection{Mixed states composed of orthogonal projections and non orthogonal
projections}

Consider a mixture
$$\rho  =p_{\Psi^{+}} \vert \Psi^{+} \rangle+ p_{\Phi^{+}} \vert \Phi^{+} \rangle
=
{1\over 2}
\left(
\begin{array}{cccc}
p_{\Psi^{+}}&0&0&p_{\Psi^{+}}\\
0&p_{\Phi^{+}}&p_{\Phi^{+}}&0\\
0&p_{\Phi^{+}}&p_{\Phi^{+}}&0\\
p_{\Psi^{+}}&0&0&p_{\Psi^{+}}
\end{array}
\right)
$$
of the two orthogonal Bell states
$\vert \Psi^{+} \rangle $
and
$\vert \Phi^{+} \rangle $, with $p_{\Psi^{+}}+p_{\Phi^{+}}=1$.
The corresponding frame function is
$
f(x)=(1/2)\left\{
p_{\Psi^{+}}(x_1+x_4)^{2} +
p_{\Phi^{+}}(x_2 +x_3)^{2}\right\}$.

The eigenvalues of the density matrix are $p_{\Psi^{+}}$ and $p_{\Phi^{+}}$, and in its diagonalized
form the matrix, is given by
$
{\rm diag }(p_{\Psi^{+}},p_{\Phi^{+}},0,0)
$.
The associated frame function is $f(x)= p_{\Psi^{+}} x_1^{2}+p_{\Phi^{+}}x_2^{2}$.

Let us now consider an example in which the projectors are not orthogonal.
$$
\begin{array}{ccc}
\rho &=& a \vert \varphi_1 \rangle +b \vert \varphi_2 \rangle ,\\
\vert \varphi_1 \rangle&=& {\rm diag} (1,0,0) ,\\
\vert \varphi_2 \rangle&=&
{1\over 2}
\left(
\begin{array}{ccc}
1&1&0\\
1&1&0\\
0&0&0
\end{array}
\right)    ,
\end{array}
$$
such that
$$
\rho ={1\over 2}
\left(
\begin{array}{ccc}
a+{b\over 2}&{b\over 2}&0\\
{b\over 2}&{b\over 2}&0\\
0&0&0
\end{array}
\right).
$$
The associated frame function is
$f(x)=a x_1^{2}+{b\over 2}(x_1+x_2)^{2}$.
Of course we can diagonalize this matrix too, and obtain
$$\rho = {\rm diag}\left(
{1\over 2}+\sqrt{{1\over 4}-{ab\over 2}},
{1\over 2}-\sqrt{{1\over 4}-{ab\over 2}},0\right),$$ so that the frame function is
$$
f(x)= \left({1\over 2}+\sqrt{{1\over 4}-{ab\over 2}}\right)x_1^{2 }+
\left({1\over 2}-\sqrt{{1\over 4}-{ab\over 2}}\right)x_2^{2}
.$$

\section{The hard part: density operators from frame functions}

The previous section dealt with the straightforward task of finding a frame function from a given quadratic form.
Now we shall deal with the less straightforward inverse problem; i.e.,
finding the quadratic form associated from a given frame function, or probability distribution.
By Gleason's theorem we know
that all frame functions correspond to positive symmetric quadratic forms.
In general, the problem can be answered by enumerating the solution of the
system of linear equations
$$f(x)=\langle x\vert \rho \vert x\rangle.$$

We present some examples.
Consider the frame function
$$f(x)=
{1\over 7}\left[3x_1^{2} +2\left(x_2-x_3\right)^{2}\right].$$
For this function the density operator has the form
$$\rho =
{1\over 7}
\left(
\begin{array}{ccc}
3&0&0\\
0&2&-2\\
0&-2&2
\end{array}
\right),
$$
which
is the sum of two orthogonal projectors
$$
\begin{array}{ccc}
E_1&=&{\rm diag}(1,0,0)\\
E_2&=&{1\over 2}
\left(
\begin{array}{ccc}
0&0&0\\
0&1&-1\\
0&-1&1
\end{array}
\right),
\end{array}
$$
with statistical weights $3/7$ and $4/7$ for $E_1$ and $E_2$, respectively.

As a nonorthogonal case, consider
the function
$$f(x)=
{1\over 12}\left[4x_1^{2}+3\left(x_1-x_2\right)^{2}+\left(x_2-x_3\right)^{2}\right].$$
Inspection shows that this function is a positive quadratic form; hence $f(x)$
is a frame function. The density operator of the system has the following
form
$$ \rho =  {1\over 12}
\left(
\begin{array}{ccc}
7&-3&0\\
-3&4&-1\\
0&-1&1
\end{array}
\right).
$$
This density operator is not generated by orthogonal states anymore.

\subsection{General classification of frame functions}

The signature of a real quadratic form  $f(y)= y^TAy$ or of a   symmetric bilinear form
is the number of positive, negative, and zero eigenvalues of the corresponding matrix $A$.
Sylvester's law of inertia states that
the signature is an invariant of the quadratic form; i.e., it is independent of the choice of basis.
Stated differently
\cite[p. 1062]{Gradshteyn},
when a quadratic form associated with the matrix $A$ is reduced by a nonsingular linear transformation $S$
such that
$A'= SAS^T$ ($S$ is a nonsingular matrix) to
$$f'(y) = y^TA'y= y_1^2 +y_2^2+ \cdots +y_P^2 - y_{P+1}^2 - y_{P+2}^2 - \cdots -y_{P+N}^2,$$
the numbers $P$ and $N$ of positive and negative squares appearing in the reduction is an invariant
of the quadratic form  and does not depend on the method of reduction.


So far we have seen that every symmetric quadratic form gives rise to a
frame function $f(x)$. By  Sylvester's law of inertia we know that we
can find a Sylvester basis for every symmetric quadratic form in which the
diagonal matrix has only positive, negative, and zero  entries.
With the probabilistic interpretation,  our
quadratic forms $\rho$ have to be positive definite, and thus the negative value  can be omitted.
Hence,
for the $n$-dimensional case, all quadratic forms, and thus all
frame functions, are isomorphic (up to a coordinate changes and permutations) to $n+1$ types of
frame functions corresponding the different signatures of the resulting density operators.

\section{Non-Gleason type probabilities}

In what follows, two examples for non-Gleason type probability measures on
quantized systems will be discussed. A common feature of such probabilities
is that they are not ``smooth'' and are based on singular distributions.

\subsection{Two-dimensional cases}

\begin{figure}[htd]
\unitlength 0.40mm
\linethickness{0.4pt}
\begin{picture}(151.06,45.00)
\put(0.00,35.00){\circle{2.11}}
\put(0.00,35.00){\circle*{3.50}}
\put(60.00,35.00){\circle{2.11}}
\put(0.00,26.37){\makebox(0,0)[cc]{$p_-^1$}}
\put(60.00,26.37){\makebox(0,0)[cc]{$p_+^1$}}
\put(0.00,35.00){\line(1,0){60.00}}
\put(70.00,35.00){\circle{2.11}}
\put(130.00,35.00){\circle{2.11}}
\put(70.00,35.00){\circle*{3.50}}
\put(70.00,26.37){\makebox(0,0)[cc]{$p_-^2$}}
\put(130.00,26.37){\makebox(0,0)[cc]{$p_+^2$}}
\put(70.00,35.00){\line(1,0){60.00}}
\put(30.00,45.00){\makebox(0,0)[cc]{$L(x^1)$}}
\put(100.00,45.00){\makebox(0,0)[cc]{$L(x^2)$}}
\put(10.00,5.00){\circle{2.11}}
\put(10.00,5.00){\circle*{3.50}}
\put(70.00,5.00){\circle{2.11}}
\put(10.00,-3.63){\makebox(0,0)[cc]{$p_-^3$}}
\put(70.00,-3.63){\makebox(0,0)[cc]{$p_+^3$}}
\put(10.00,5.00){\line(1,0){60.00}}
\put(90.00,5.00){\circle{2.11}}
\put(90.00,5.00){\circle*{3.50}}
\put(150.00,5.00){\circle{2.11}}
\put(90.00,-3.63){\makebox(0,0)[cc]{$p_-^n$}}
\put(150.00,-3.63){\makebox(0,0)[cc]{$p_+^n$}}
\put(90.00,5.00){\line(1,0){60.00}}
\put(40.00,15.00){\makebox(0,0)[cc]{$L(x^3)$}}
\put(120.00,15.00){\makebox(0,0)[cc]{$L(x^n)$}}
\put(80.00,4.67){\makebox(0,0)[cc]{$\cdots$}}
\end{picture}
\\
(a)\\
$ $\\
\unitlength 0.40mm
\linethickness{0.4pt}
\begin{picture}(151.06,45.00)
\put(0.00,33.00){\rule{4\unitlength}{4\unitlength}}
\put(60.00,33.00){\rule{4\unitlength}{4\unitlength}}
\put(0.00,26.37){\makebox(0,0)[cc]{$p_-^1$}}
\put(60.00,26.37){\makebox(0,0)[cc]{$p_+^1$}}
\put(0.00,35.00){\line(1,0){60.00}}
\put(70.00,33.00){\rule{4\unitlength}{4\unitlength}}
\put(130.00,33.00){\rule{4\unitlength}{4\unitlength}}
\put(70.00,26.37){\makebox(0,0)[cc]{$p_-^2$}}
\put(130.00,26.37){\makebox(0,0)[cc]{$p_+^2$}}
\put(70.00,35.00){\line(1,0){60.00}}
\put(30.00,45.00){\makebox(0,0)[cc]{$L(x^1)$}}
\put(100.00,45.00){\makebox(0,0)[cc]{$L(x^2)$}}
\put(10.00,3.00){\rule{4\unitlength}{4\unitlength}}
\put(70.00,3.00){\rule{4\unitlength}{4\unitlength}}
\put(10.00,-3.63){\makebox(0,0)[cc]{$p_-^3$}}
\put(70.00,-3.63){\makebox(0,0)[cc]{$p_+^3$}}
\put(10.00,5.00){\line(1,0){60.00}}
\put(90.00,3.00){\rule{4\unitlength}{4\unitlength}}
\put(150.00,3.00){\rule{4\unitlength}{4\unitlength}}
\put(90.00,-3.63){\makebox(0,0)[cc]{$p_-^n$}}
\put(150.00,-3.63){\makebox(0,0)[cc]{$p_+^n$}}
\put(90.00,5.00){\line(1,0){60.00}}
\put(40.00,15.00){\makebox(0,0)[cc]{$L(x^3)$}}
\put(120.00,15.00){\makebox(0,0)[cc]{$L(x^n)$}}
\put(80.00,4.67){\makebox(0,0)[cc]{$\cdots$}}
\end{picture}
\\
(b)
\caption{\label{f-gd-monm}
Examples for a non-Gleason type probability measure
for $n$ spin one-half state propositional systems $L(x^i),
i=1,\cdots
,n$
which are not comeasurable. The superscript $i$ represents the $i$th
measurement direction.
(a) Full circles indicate the atoms with probability measure 1.
There exists a classical but no quantum realization of this state.
(b) Full squares indicate atoms with probability measure 1/2.
The quantum realization is the ``most ignorant state.''
}
\end{figure}
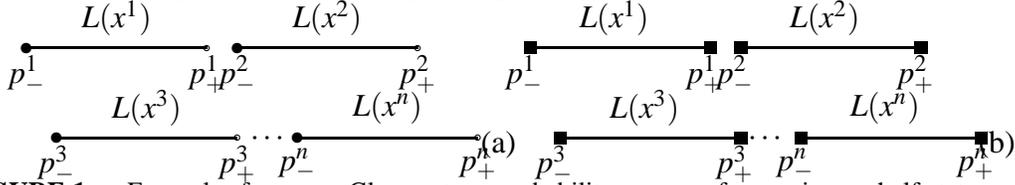

As in two dimensions, single lattices are not tied
together at common basis elemens, there exists the possibility of non-Gleason type probability
measures; i.e., measures which have singular, separating distributions
and thus can be embedded into ``classical'' Boolean algebras.
One particular example \cite{svozil-ql}, depicted as Greechie diagram, in which orthogonal
bases are represented by smooth lines, connecting the basis elements
which are represented by circles.
is drawn in
Figure \ref{f-gd-monm}(a).
Its probability measure is $P(x_-^i)=1$ and $P(x_+^i)=1-P(x_-^i)=0$ for
$i=1,\ldots ,n$.
There is no quantum realization for this classic probability measure;
yet it cannot be excluded by Gleason's theorem.
Whether these non-Gleason type states are pure artifacts of
two-dimensional
Hilbert spaces or have a physical meaning remains an open problem.
One indication that such state is impossible to attain quantum mechanically is
a principle \cite{zeil-99,DonSvo01,zeil-bruk-02} stating that an elementary quantum mechanical system
in two-dimensional Hilbert space can carry only one classical bit.
The configuration drawn in Figure \ref{f-gd-monm}(b) has a quantum realization by the
state $\rho = {\rm diag}(1/2,1/2)$ corresponding to the null information.

\subsection{Three-dimensional cases}

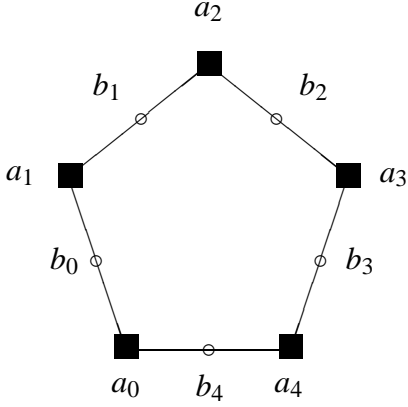
\begin{figure}
\unitlength 0.50mm
\linethickness{0.4pt}
\begin{picture}(99.67,100.00)
\put(28.33,10.33){\circle{2.75}}
\put(72.00,10.33){\circle{2.75}}
\put(50.33,86.00){\circle{2.75}}
\put(13.33,56.33){\circle{2.75}}
\put(87.33,56.33){\circle{2.75}}
\put(13.33,56.33){\line(5,4){37.00}}
\put(50.33,85.93){\line(5,-4){37.00}}
\put(13.00,56.00){\line(1,-3){15.33}}
\put(87.33,56.00){\line(-1,-3){15.33}}
\put(72.00,10.00){\line(-1,0){44.00}}
\put(20.33,33.66){\circle{2.75}}
\put(80.00,33.66){\circle{2.75}}
\put(68.33,71.33){\circle{2.75}}
\put(32.33,71.33){\circle{2.75}}
\put(50.33,10.00){\circle{2.75}}
\put(25.33,7.67){\rule{6.33\unitlength}{6.33\unitlength}}
\put(10.33,53.33){\rule{6.33\unitlength}{6.33\unitlength}}
\put(47.33,83.00){\rule{6.33\unitlength}{6.33\unitlength}}
\put(84.33,53.33){\rule{6.33\unitlength}{6.33\unitlength}}
\put(69.00,7.67){\rule{6.33\unitlength}{6.33\unitlength}}
\put(28.33,0.00){\makebox(0,0)[cc]{$a_0$}}
\put(50.67,0.00){\makebox(0,0)[cc]{$b_4$}}
\put(12.00,33.67){\makebox(0,0)[cc]{$b_0$}}
\put(0.33,56.33){\makebox(0,0)[cc]{$a_1$}}
\put(23.33,80.00){\makebox(0,0)[cc]{$b_1$}}
\put(50.33,100.00){\makebox(0,0)[cc]{$a_2$}}
\put(78.33,80.00){\makebox(0,0)[cc]{$b_2$}}
\put(99.67,56.00){\makebox(0,0)[cc]{$a_3$}}
\put(90.67,33.67){\makebox(0,0)[cc]{$b_3$}}
\put(71.67,0.00){\makebox(0,0)[cc]{$a_4$}}
\end{picture}
\caption{\label{f-lwpen}Greechie diagram of the Wright
pentagon
\protect\cite{wright:pent}. Filled squares indicate
probability~${1\over 2}$.}
\end{figure}

Another example of a suborthoposet which is embeddable into the
three-dimensional real Hilbert lattice
${ C}({\bf R}^3)$ has been given by Wright \cite{wright:pent}. Its
Greechie diagram of the pentagonal form is drawn in Figure
\ref{f-lwpen}.
An explicit embedding is \cite{svozil-tkadlec}
\begin{eqnarray*}
&&a_{ 0} ={\rm Span}( \sqrt{  \sqrt5} , \sqrt{ 2+\sqrt5} , \sqrt{
3+\sqrt5}),\\
&&b_{0} ={\rm Span}( \sqrt{  \sqrt5} ,-\sqrt{ 2+\sqrt5} , \sqrt{ 3-\sqrt5}),\\
&&a_{ 1} ={\rm Span}(-\sqrt{  \sqrt5} ,-\sqrt{-2+\sqrt5} , \sqrt  2
),\\
&&b_{1} ={\rm Span}( 0               , \sqrt 2          , \sqrt{-2+\sqrt5}),\\
&&a_{2} ={\rm Span}( \sqrt{  \sqrt5} ,-\sqrt{-2+\sqrt5} , \sqrt  2        ),\\
&&b_{2} ={\rm Span}(-\sqrt{  \sqrt5} ,-\sqrt{ 2+\sqrt5} , \sqrt{3-\sqrt5}),\\
&&a_{3} ={\rm Span}(-\sqrt{  \sqrt5} , \sqrt{ 2+\sqrt5} , \sqrt{ 3+\sqrt5}),\\
&&b_{3} ={\rm Span}( \sqrt{5+\sqrt5} , \sqrt{ 3-\sqrt5} ,2\sqrt{-2+\sqrt5}),\\
&&a_{4} ={\rm Span}( 0               ,-\sqrt{-1+\sqrt5} , 1               ),\\
&&b_{4} ={\rm Span}(-\sqrt{5+\sqrt5} , \sqrt{ 3-\sqrt5} ,2\sqrt{-2+\sqrt5}).
\end{eqnarray*}
Wright showed that the probability measure
$$P(a_i)={1\over 2},\; P(b_i)=0,\quad i=1,2,3,4$$
as depicted in Figure \ref {f-lwpen} is no convex combination of other
pure states.
Note that these configuration corresponds to a sublattice
(but not to a subalgebra) of the full Hilbert lattice of quantum propositions.
Sublattices may always allow more states than the full lattice.
Therefore, the above probability measure cannot be expected to be consistently
extendable to the entire Hilbert lattice.

\section{Summary}

We propose to put Gleason's theorem to constructive use in deriving quantum states
from frame functions indentified as probability distributions.
Thus, there exists a direct, constructive route from experimental frequency counts to
the quantum states compatible with these counts.
In this interpretation, Gleason's theorem presents an inductive element.

Since in two-dimensional Hilbert space there is no possible joint connection
of one basis with any other basis, there is no structure assuring the continuity
supporting a Gleason-type argument.
This is the reason for the impossibility to base the Born rule
on the assumptions made by Gleason.
From three dimensions onwards, bases can be interlinked in one or more common leg(s);
and systems of bases can be cyclic, as exploited in the Kochen-Specker proof.
It is not unreasonable to spaculate that,
as for two-dimensional Hilbert space there is no strong support of the Born rule
by Gleason's theorem,
it would be very interesting to find any empirical evidence for its violation.


\begin{thebibliography}{17}
\expandafter\ifx\csname natexlab\endcsname\relax\def\natexlab#1{#1}\fi
\providecommand{\enquote}[1]{``#1''}
\expandafter\ifx\csname url\endcsname\relax
  \def\url#1{\texttt{#1}}\fi
\expandafter\ifx\csname urlprefix\endcsname\relax\def\urlprefix{URL }\fi
\providecommand{\eprint}[2][]{\url{#2}}

\bibitem[von Neumann(1932)]{v-neumann-49}
J.~von Neumann, \emph{Mathematische Grundlagen der Quantenmechanik}, Springer,
  Berlin, 1932, {E}nglish translation in \cite{v-neumann-55}.

\bibitem[Gleason(1957)]{Gleason}
A.~M. Gleason, \emph{Journal of Mathematics and Mechanics} \textbf{6}, 885--893
  (1957).

\bibitem[Dvure{\v{c}}enskij(1993)]{r:dvur-93}
A.~Dvure{\v{c}}enskij, \emph{{G}leason's Theorem and Its Applications}, Kluwer
  Academic Publishers, Dordrecht, 1993.

\bibitem[Cooke et~al.(1985)]{c-k-m}
R.~Cooke, M.~Keane, and W.~Moran, \emph{Math. Proc. Camb. Soc.} \textbf{98},
  117--128 (1985).

\bibitem[Peres(1993)]{peres}
A.~Peres, \emph{Quantum Theory: Concepts and Methods}, Kluwer Academic
  Publishers, Dordrecht, 1993.

\bibitem[Hrushovski and Pitowsky(1989)]{hru-pit-2003}
E.~Hrushovski, and I.~Pitowsky, Generalizations of {K}ochen and {S}pecker's
  theorem and the effectiveness of {G}leason's theorem (1989), preprint,
  \eprint{quant-ph/0307139}.

\bibitem[Richman and Bridges(1999)]{rich-bridge}
F.~Richman, and D.~Bridges, \emph{Journal of Functional Analysis} \textbf{162},
  287--312 (1999), \urlprefix\url{http://dx.doi.org/10.1006/jfan.1998.3372}.

\bibitem[Kochen and Specker(1967)]{kochen1}
S.~Kochen, and E.~P. Specker, \emph{Journal of Mathematics and Mechanics}
  \textbf{17}, 59--87 (1967), reprinted in \cite[pp. 235--263]{specker-ges}.

\bibitem[Gradshteyn and Ryzhik(2000)]{Gradshteyn}
I.~S. Gradshteyn, and I.~M. Ryzhik, \emph{Tables of Integrals, Series, and
  Products, 6th ed.}, Academic Press, San Diego, CA, 2000.

\bibitem[Svozil(1998)]{svozil-ql}
K.~Svozil, \emph{Quantum Logic}, Springer, Singapore, 1998.

\bibitem[Zeilinger(1999)]{zeil-99}
A.~Zeilinger, \emph{Foundations of Physics} \textbf{29}, 631--643 (1999).

\bibitem[Donath and Svozil(2002)]{DonSvo01}
N.~Donath, and K.~Svozil, \emph{Physical Review A} \textbf{65}, 044302 (2002),
  \urlprefix\url{http://dx.doi.org/10.1103/PhysRevA.65.044302},
  \eprint{quant-ph/0105046}.

\bibitem[Brukner and Zeilinger(2003)]{zeil-bruk-02}
{\v{C}}.~Brukner, and A.~Zeilinger, \enquote{Information and fundamental
  elements of the structure of quantum theory,} in \emph{Time, Quantum and
  Information}, edited by L.~Castell, and O.~Ischebek, Springer, Berlin, 2003,
  pp. 323--355, \eprint{quant-ph/0212084}.

\bibitem[Wright(1978)]{wright:pent}
R.~Wright, \enquote{The state of the pentagon. {A} nonclassical example,} in
  \emph{Mathematical Foundations of Quantum Theory}, edited by A.~R. Marlow,
  Academic Press, New York, 1978, pp. 255--274.

\bibitem[Svozil and Tkadlec(1996)]{svozil-tkadlec}
K.~Svozil, and J.~Tkadlec, \emph{Journal of Mathematical Physics} \textbf{37},
  5380--5401 (1996), \urlprefix\url{http://dx.doi.org/10.1063/1.531710}.

\bibitem[von Neumann(1955)]{v-neumann-55}
J.~von Neumann, \emph{Mathematical Foundations of Quantum Mechanics}, Princeton
  University Press, Princeton, 1955.

\bibitem[Specker(1990)]{specker-ges}
E.~Specker, \emph{Selecta}, Birkh{\"{a}}user Verlag, Basel, 1990.

\end{thebibliography}

\end{document}